\documentclass[sigplan,11pt,nonacm=true]{acmart}
\AtBeginDocument{%
  \providecommand\BibTeX{{%
    \normalfont B\kern-0.5em{\scshape i\kern-0.25em b}\kern-0.8em\TeX}}}

\setcopyright{acmcopyright}
\copyrightyear{2018}
\acmYear{2018}
\acmDOI{XXXXXXX.XXXXXXX}

\acmConference[Conference acronym 'XX]{Make sure to enter the correct
  conference title from your rights confirmation emai}{June 03--05,
  2018}{Woodstock, NY}
\acmPrice{15.00}
\acmISBN{978-1-4503-XXXX-X/18/06}

\begin{document}

\title{Applying Machine Learning Analysis for Software Quality Test}

\author{Al Khan}
\email{pkhan_1@hotmail.com}
\author{Remudin Reshid Mekuria}
\email{remudin@alatoo.edu.kg}
\author{Ruslan Isaev}
\email{ruslan.isaev@alatoo.edu.kg}
\affiliation{
  \institution{Ala-Too International University}
\city{Bishkek}
  \country{Kyrgyzstan}
}


\begin{abstract}
One of the biggest expense in software development is the maintenance. Therefore, it’s critical to comprehend what triggers maintenance and if it may be predicted. Numerous research outputs have demonstrated that specific methods of assessing the complexity of created programs may produce useful prediction models to ascertain the possibility of maintenance due to software failures. As a routine it is performed prior to the release, and setting up the models frequently calls for certain, object-oriented software measurements. It’s not always the case that software developers have access to these measurements. In this paper, machine learning is applied on the available data to calculate the cumulative software failure levels. A technique to forecast a software’s residual defectiveness using machine learning can be looked into as a solution to the challenge of predicting residual flaws. Software metrics and defect data were separated out of the static source code repository. Static code is used to create software metrics, and reported bugs in the repository are used to gather defect information. By using a correlation method, metrics that had no connection to the defect data were removed. This makes it possible to analyze all the data without pausing the programming process. Large, sophisticated software’s primary issue is that it is impossible to control everything manually, and the cost of an error can be quite expensive. Developers may miss errors during testing as a consequence, which will raise maintenance costs. Finding a method to accurately forecast software defects is the overall objective.

\end{abstract}



\maketitle

\section{Introduction}

A software feature is a functional component \cite{Sevn}. Creating software is frequently centered on creating features. Development of features is carried out such that one or more features are not created by several teams simultaneously. Errors in program are frequently unevenly distributed throughout all of its components or features \cite{stuart2020applying}. Instead, certain features have a tendency to fail more frequently than others. The next issue is whether it is possible to detect these errors before they start to result in failures. If so, this may make it possible to examine these characteristics more closely before implementing for the clients. It is a general accepted fact that defects are unavoidable when designing large-scale software. Therefore, it is crucial for when creating software to have a way to address flaws as they evolve into failures. Failures that aren't managed well can cost a lot of money and damage your reputation with customers. While proactive handling in particular and competent failure management in general may cut costs and boost customer trust. 
How may the repercussions of software failures be decreased? For this, two strategies are taken into account. First, determine how high a release's failure rate is expected to be. And second, by narrowing the focus to identify which software features are more prone to errors in a future release. This demands for a system that can forecast impending failures and predict developed software features according to how likely they are to lead to mistakes. Such a system ought to be impartial, clear, and accurate (to the extent that is possible and feasible). This is necessary because it must be simple to compare outcomes across the organization, simple to deploy, and useful.
It may be argued that machine learning methods are suitable to defectiveness prediction tasks based on common methodologies. To date, there are now more research demonstrating machine learning algorithms for software maintenance than other methods. There is no single and or unique algorithm that can be used to all static software \cite{wolpert1997no}. Because of this, it is often hard to utilize the same prediction model across all software. Research methodologies that are more trustworthy should be created to ensure that all comparative studies are correct \cite{myrtveit2005reliability}.

\begin{figure}[h!]
    \centering
    \includegraphics[scale=0.5]{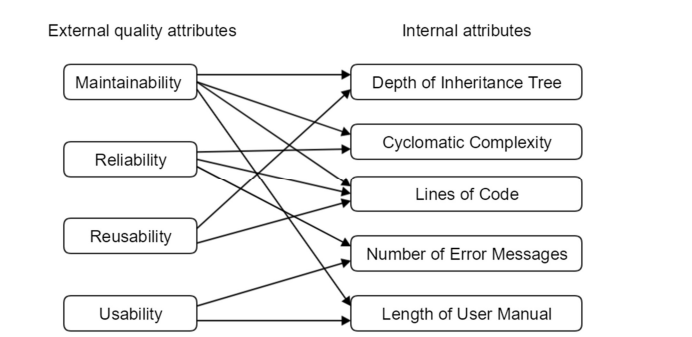}
    \caption{Types of software metrics.}
    \label{fig:fig1}
\end{figure}

The frequency of errors reduces with decreasing software unit size, according to a research \cite{basili1984software}, that examined the relationship between software unit size and defect quantity. However, other studies \cite{hatton1997reexamining} also demonstrate that the quantity of errors is connected to the so-called "ideal size," which is neither endlessly big or is of the smallest size. Two studies \cite{moller1993empirical, neil1992multivariate} explain why there is no clear correlation between the size of a software unit and the quantity or density of flaws in it from the third angle. The main goals are to identify new dependencies in the qualities of the code itself, to develop prediction models for software (identifying flaws as the program is being developed to help identify future restrictions), and to attempt to apply this knowledge to other software projects. This research's new findings should make it easier to comprehend present software issues and anticipate failures in the future.
Generally speaking, the use of metrics enables project and enterprise managers to analyze the complexity of a project that is being developed or even one that has already been developed, as well as the scope of work, the style of the program being developed, and the efforts each developer expended to implement a specific solution. Metrics, however, can only be used as advisory traits; they cannot be entirely guiding, as programmers often want to minimize or maximize a measure for their program, which can lead to the use of methods that reduce program performance. Additionally, if a programmer just added a few lines of code or made a few structural modifications, for instance, this does not necessarily mean that he accomplished nothing; rather, it may indicate that the program's flaw was exceedingly difficult to detect. However, the latter issue may be somewhat resolved by employing complexity metrics because a more complicated program makes it harder to discover the fault.     Importantly, there is no specific general metric that applies to all situations. Hybrid measures can be used, but they also rely on simpler metrics and cannot be universal. Any controllable metric aspects of the program must be regulated either relying on each other or depending on the individual job. No measure can be incorporated into the absolute and judgments made solely based on it since, strictly speaking, any metric is only an indicator that much depends on the language and programming style of the program. Figure \ref{fig:fig1} gives an illustration of how qualities relate to one another. On the basis of a certain set of internal qualities, the quality attributes in this figure are computed (basic software metrics). Metrics are extracted, and then they are examined to look for any relevant relationships. The investigation of really existent relationships between measures is the main purpose of data analysis. Identification of the cause-effect correlations between the indicators, or how much the change in one indicator depends on the change in another, is important for conducting a relational research of these links.

\begin{figure}[h!]
    \centering
    \includegraphics[scale=0.5]{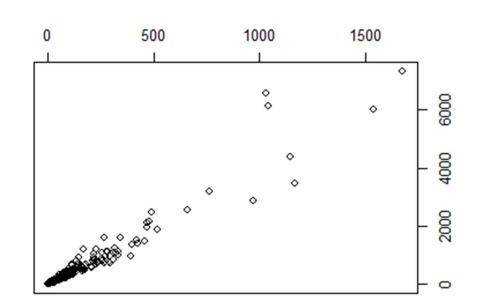}
    \caption{Software metrics Weighted Method Count (WMC) as X-axis and Lines of Code (LOC) as Y-axis scatter.}
    \label{fig:fig2}
\end{figure}

The influence of individual components may only be seen as a tendency (on average) in the mass observation of real data, which is how correlations may be related. Examples of correlation dependency include the link between a bank's asset size and profit, the rise in labor productivity, and the number of years a person has worked there. Pair correlation, or the link between two attributes, is the most basic form of correlation dependency (productive and factorial or between two factorial). This dependency may be mathematically stated as the relationship between the effective exponent y and the factor exponent x. Direct and reverse connections (graph edges) are also possible. In the first scenario, as the sign of x grows, so does the sign of y; conversely, as the sign of x lowers, so does the sign of y. Figure \ref{fig:fig2} displays an example of a metrics scatter plot

\section{Literature Review}

There are several approaches available for defect prediction in software projects, which has been the subject of much study \cite{rana2014defect}. This is based on anticipating the overall flow of a project or identifying the components that are most likely to produce problems. According to the 80-20 rule, 20\% of the features in software will have 80\% of the problems. Knowing that is only helpful if you are familiar with the capabilities of that software. Software feature clustering is a related topic that, as a result, has produced results that have mostly been used to various reuse initiatives \cite{srinivas2015feature}.
The selection of prediction measure employed in fault prediction is crucial \cite{shihab2013lines}. There are three primary categories of metrics:
\begin{itemize}
    \item Object oriented metrics
    \item Non-object oriented metrics
    \item Process metrics
\end{itemize}

First and foremost, the length, size, complexity, and cohesiveness of the generated program are measured by the first two prediction metrics \cite{staron2018software}. The latter evaluates the method used to construct the software. Object-oriented metrics are concerned with object-oriented programming and the organization of the developed applications. Examples of possible metrics are the quantity of public metrics, inheritances, couplings, or methods per class \cite{singh2018taxonomy}.
Non-object oriented metrics have to do with things like complexity or size in terms of lines of code. For instance, the MacCabe complexity measure, which assesses independent program flow pathways, quantifies program complexity \cite{maccabe1983structured}. Different change metrics, including revisions, bugfixes, age, and others, are related to process measurements. These change measurements have been shown to produce reliable fault prediction outcomes \cite{moser2008comparative}. Accordingly, object-oriented measures frequently correlate with non-object-oriented metrics and have a tendency to provide the best performance \cite{shihab2013lines}. There is a link between them even if object-oriented measures, such size or complexity metrics, collect more data than non-object-oriented ones. This is expected since active lines of code always increase complexity as larger structures of any type tend to become more complicated \cite{staron2018software}. Although complexity measurements are meant to better reflect the nature of a program, it has been demonstrated that they can, at least in certain circumstances, be worse than straightforward approaches like lines of code \cite{menzies2002metrics}.
The significance of selecting the right metric was demonstrated in a study where a substantial association between inheritance metrics and the quantity of bugs in particular software components was discovered \cite{cartwright2000empirical}. According to the study, classes with inheritance structures were around three times more likely to have errors than classes without them. Additionally, classes lower in the inheritance chain seemed to be more likely to have flaws. They were able to build viable prediction models by using this measure together with other object-oriented metrics like tracking events and states per class.
The object-oriented method has been used by others to demonstrate that failure-proneness may be predicted relatively early on in a project \cite{kumar2017empirical}. Its restriction to object-oriented programming is a limitation. However, since size and complexity measurements do correlate with object-oriented metrics, it is likely that they are also helpful.
Because complexity and size are connected. The fact that lines of code (LOC) have been effectively applied as a defect predictor \cite{zhang2009investigation} may not come as a surprise. This has been demonstrated to improve when additional process measurements are included \cite{ostrand2005predicting}. Increasing the likelihood that a combination would be the best choice if both process measurements and non-object oriented metrics are available.
This study focuses on cumulative failure-prediction, which seeks to estimate the total number of reported problems in a software deployment and to look at if specific kinds of software features are more prone to errors. Two domains in the defect prediction field that somewhat match to these objectives are defect inflow prediction \cite{staron2008predicting} and software error prediction \cite{rana2014defect}. The former focuses on a specific product with changing levels of granularity, whereas, the latter is concerned with an overall amount of defect ingress. According to research on forecasting defect backlog, outcomes must be simple to comprehend in order to be useful.  According to the report, development strategy must need knowledge on the overall predicted trajectory of the defect backlog and course dependability. As a result, more esoteric and complex models may become less valuable as the information they provide gets more complicated, maybe more accurate, but also more difficult to utilize \cite{staron2010method}.
Through the use of multivariate linear regression techniques \cite{staron2008predicting}, which have been shown to have an accuracy of within 72\% for a forecast of 3 weeks when using traits from work packages that are currently being developed. Two extremely basic regressive approaches depending can be created \cite{fenton1999critique} on either lines of code or complexity in another investigation. But after that, need to provide Bayesian Belief Nets for predicting failure. In addition to program, size, and some level of complexity data, the regression analysis has been further refined to make predictions in test environments \cite{suffian2014prediction}.
To forecast the inflow of defects in an ongoing project, a different research employed Bayesian inference \cite{rana2016analyzing}. This method produced better results than many others, including multivariate linear regression, expert judgment, and analogy-based estimation. The major issue is that the distribution needs to be estimated, which, depending on the data, might be problematic. The error was between 12\% and 21\%. Is the forecasting of the propensity for bugs in certain software components. The strategies employed will determine how precise these forecasts are \cite{rana2014defect}. Among other options when employing the machine learning technique, tree-based methods like random forests, neural networks, and support vector machines are frequently used \cite{li2019evaluating}. The first two's accuracy was demonstrated to be pretty excellent, but the payoff is questionable due to the intricacy of their parameter-tuning and training. Some research used object-oriented metrics to predict release failure in Eclipse post-release, and the results were computed using univariate binary regression and univariate multiple regression. 

This was done in order to foresee the number of failures as well as the various degrees of severity of those failures \cite{shatnawi2008effectiveness}. The study also demonstrated that following a release, predictions are harder. This may be because there has been less work and fewer failures have occurred as a result of testing before release. Another demonstrates that once an appropriate measure had been selected, Neural Networks worked effectively in predicting failure-prone components of a program \cite{kumar2017empirical}. A thorough analysis of machine learning methods revealed that they frequently beat linear regression models. The analysis revealed that Random Forests were frequently the most effective models \cite{malhotra2015systematic}. This is most likely caused by the algorithm's noise resistance and tweaking capabilities, as demonstrated by Guo et al. \cite{guo2004robust}. Though additional measures should also be tested, their study mainly employed MacCabe measurements. Programmers have utilized clustering to group software features \cite{shah2013software}. The results show that it is more challenging when there are no labels provided. Additionally, automatic clustering frequently performs worse than professionals. However, the last one shouldn't come as a surprise. Accordingly, the degree of clustering can significantly affect performance.
Software clustering is typically employed when ground-truth data on prior failures is available. However, attempts to anticipate errors without access to underlying data have also been made using software clustering \cite{bishnu2011software}. With the use of a method to identify the cluster centers prior to initializing the k-means algorithm, it was demonstrated that this was feasible.

\section{Research Methodology}

The purpose of this research is to investigate the possibility to combine software clustering with fault-prediction. This created two main research questions, as outlined below, with related sub-questions.
\newline
\textbf{RQ 1:} How to predict the number of failures in a software using Machine Learning?
To reduce the number of algorithms examined, the algorithms that will be put to the test are chosen in advance. The main focus of this research is repeated in the most recent version with new input data as it becomes available over the course of the project.
\newline
\textbf{RQ 2:} How can clusters of software features be created using machine learning to aid in failure prediction?
The goal is to enhance the dataset utilized by incorporating aggregated data on the established characteristics once the basic prediction algorithms have been understood in order to presumably improve performance. This question is evaluated based on the silhouette score as well as the cluster distribution, and clustering that would produce extremely sparse input by test three was deemed undesirable because it could imply that the input is not being taken into account at all, particularly in regression problems.
\newline
\textbf{RQ 2.1:} What data regarding the software features is best used for clustering?
This is as a result of the wider issue of the dimensionality. However, due to this the data needs to be reduced and a balance maintained. In general, more data and variables equal greater information. 
\newline
\textbf{RQ 2.2:} Are certain clusters more likely to contribute to faults in the software, and if so can they be used to increase performance of the prediction models?
Combining the prediction models from question 1 and the clustering from question 2, this question should also provide an explanation of the potential causes of a certain software feature to be problematic. When it comes to things like feature size and development time, this is now primarily done within the organization by intuition and guesswork thus bringing up another question 2.3
Answering these inquiries should accomplish the goal of this paper, which is to forecast when software problems will occur and identify the traits that are most likely to result in these failures.

\subsection{Research Design}

The definition of machine learning is "computational approaches employing experience to enhance performance or to generate accurate predictions” \cite{Mehryar2012}. Experience is the previous knowledge that is currently available, typically in the form of data. There are several applications of machine learning in each of the proposed solutions to the issue under examination. Clustering and supervised learning, which are the two types employed, are described below. Additionally, each kind has a unique algorithm with unique strengths and weaknesses. The algorithms that were thought to be solutions are also covered in detail in their respective sections. Finally, scalers, which structure data so that algorithms may use it, and metrics, which score and contrast the output of algorithms are discussed. Supervised Learning
In general, machine learning predicts some output based on input. Specifically for supervised learning, some of the input is labelled, meaning that it has a predefined output \cite{hastie2009elements}. Calling the input $x$, and the connected output $y$, we can then model the connection between the input and output as some function $f$, such that $ y = f(x) $. A supervised learning algorithm then, is a function $f'(x,\theta)$ that approximates $f$: $f'(x,\theta) \sim f(x) $,where $\theta$ is some parameter(s) in the algorithm. During the training phase, which is specific to supervised learning, $\theta$  can be updated based on the accuracy of the approximating function. This updating is done to try and improve the model performance and is based on an update policy, and the error between proposed output $y'$, calculated by the algorithm, and the true output $y$. When the training phase is concluded the objective of the algorithm is then to make predictions without this feedback on its performance as it is now possible to use the algorithm on unlabeled data, meaning that, for an input $x$ it is not necessarily the case that a connected output $y$ is known. The error, $e$, for one output-input pair which can be used as a performance measure for one update is:
\begin{equation}
    e = ||y' - y||	
\end{equation}

Where $|| \ \cdot \ ||$ signifies a norm, normally $`1$ or $`2$ 

There are two types of supervised learning problems considered in this work, classifier and regressor problems. A classifier is a type of problem aiming to predict nominal data, which belongs to certain group without internal order such as blue or grey. Or ordinal data, which is grouped data with some sort of order, such as small, medium, large. A classifier algorithm then is concerned with classification of some data set into classes \cite{sammut2017encyclopedia}.
A regression problem is that of finding a function which conveys the relationship between two or more variables. This can for instance be done when given a set of related points of x and y. To estimate a function between the two sets is a regression problem. The estimated function is sometimes, referred to as the fitted line. Regressors are algorithms which solve these problems of estimating functions.

\subsection{The Design Cycle}
\begin{figure}[h!]
    \centering
    \includegraphics[scale=0.2]{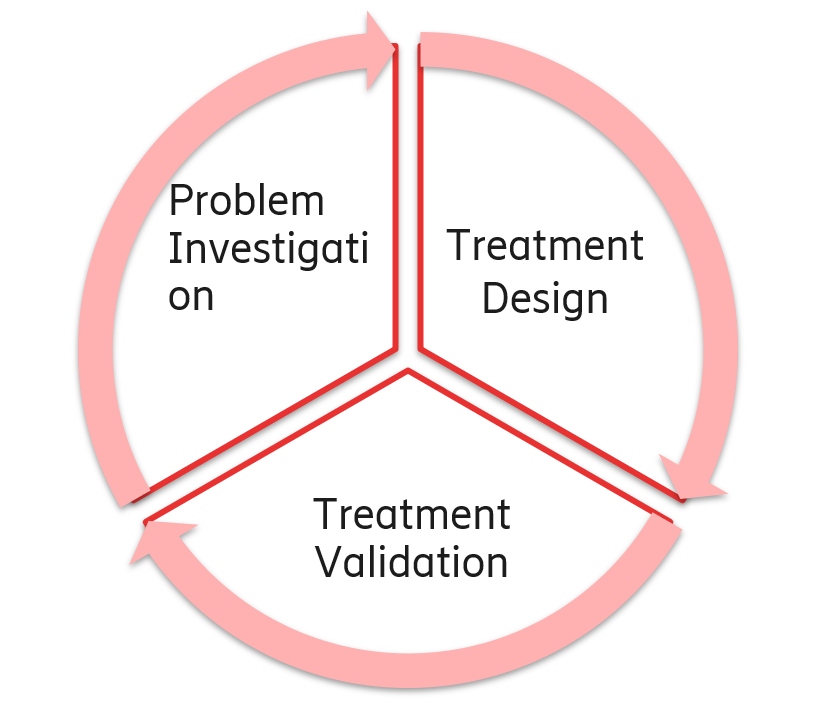}
    \caption{The Design Cycle.}
    \label{fig:fig3}
\end{figure}

Each test  follows the design cycle that Wieringa, R.J. \cite{scikit-learn} proposed and is illustrated in Figure \ref{fig:fig3}. Wieringa does not include treatment implementation in the design cycle, hence it is a condensed form of the engineering cycle. The process of comprehending both the nature of the problem and the interests of the team is outlined in problem inquiry. This prepares for treatment design, which refers to knowing the precise needs and identifying which of those aid in goal achievement. This involves identifying existing therapies for the issues and developing new ones. The process's last phase involves attempting to assess whether the therapies are appropriate for the problem and context and are capable of having the desired outcomes. The procedure is typically repeated, which is why it is depicted as a circle.

\subsubsection{Data set}
The complete data collection was composed of information that was collected ten years ago. Since the corporation changed its release strategy at that time, only the most recent four years of data were used. In the middle of 2015, they switched from biannual to monthly releases of new software features.

The aggregated failures in the software were significantly affected, and it was evident that this may have disturbed the models. After switching to monthly, the average number of failures each week was less than half. In this way, it is comparable to training the models on two totally independent sets of data with different underlying structures. The data set used was made up of four subsets:
\begin{itemize}
    \item Trouble reports
    \item Software feature data
    \item Commit data
    \item Miscellaneous information
\end{itemize}
   
Release date lists and other information that may be utilized to connect the other data sets were regarded as supplemental data. For instance, this assisted in connecting the commits that made up a software feature by comparing the date of the most recent commit to the software that would be the next one available. Failures might have been uncovered internally, through testing, or externally, by a client, and were described in trouble reports. Each trouble report provided information on the date, version, and severity of the failure. Even though it offered more information, what was used was what had already been covered. The software that were examined were associated with about 6,000 produced defect reports of varying severity. Before the features were ever made public, internal testing found the majority of the flaws. Software feature data includes high level details about the features, the date it was finished, and the period of development. Although all features had IDs to identify them, as will be seen in the example below, these IDs were distinct from those used to identify the features in the commit data subset. In all versions that were examined, there were little over 2,000 software features. The commit data contained a description of each commit made to the main branch. Number of lines of code added, removed, or changed specifically, where in the system in which changes were made of which software feature those changes were a part of. Depending on how the development teams work, modifying a small section of the system or it might include modifications to multiple separate subsystems. Each commit was placed in a group with others that had the same ID. All of the software features had a total of 19 subsystems.

Each subsystem that appeared fewer than 13 times across the whole data set was consolidated into a single group called Single Group.

\subsection{Test 1}
In this Test, the goal was to use fundamental facts to attempt a solution to research question 1. This was done to aid in understanding the issue, the data that was available, and the method that was most likely to produce positive outcomes. In this case, four supervised learning algorithms were chosen, with one left to be chosen at a later time for test three. Particularly, the fundamental data omitted additional in-depth information about the program characteristics. 
\subsection{Test 2}
This test comprised feature extraction and selection in order to address question 2. Test 2's more thorough queries were as follows:

\textbf{    a- What scalers applied?}

The scalers that were utilized were MinMax, Robust and Quantile Transformer. These were picked to illustrate several alternatives and different levels of intrusion into the original data structure. A clustering algorithm's results can be impacted by scaling since scalers change the distance between points, and clustering algorithms rely on this distance to function. The performance of the scalers was assessed once more using the silhouette score.

\textbf{    b- Which algorithms?}
    
The algorithms DBSCAN and k-means were employed. Given the widespread usage of k-means, omission would require compelling justification based on the nature of the issue. DBSCAN was selected since it should be able to handle the expected data noise effectively.
The final version sought to respond to research questions 2 and 2.1.1. The objective is to increase the data input into the models created in test 1 in order to enhance them and determine whether any software features exist that are more likely to correlate with or contribute to the reported number of failures. Trials are done with part of the clusters deleted to see whether this impacts model performance in order to better understand which clusters have a stronger impact on the inflow of defects. It's possible that clusters are less likely to result in failures if they can be eliminated without impairing the model's performance. One of the biggest issues with test 2 was how utilizing software as a time unit severely constrained the number of datapoints accessible. Some of the models were modified to account for this noise in an effort to improve the data. Grid search was utilized to fine-tune parameters like dropout and neurons in the layers in order to improve ANN performance.

\section{Results }
\subsection{Test 1 Setup}
In the initial test, RQ 1 was the focus. Aggregated information on the most recent software features was used for this. Mean development time for the features included in the software, the quantity of features that were made available, and the volume of reported failures that had previously been received. It was decided to use a latency of t = 4 after conducting tests with a bigger lag. However, the algorithms that can consistently rate their input place information from earlier times below the top 10. The algorithms, and their parameters are presented below:

\textbf{LSTM}

The two LSTM networks are built up as shown in Tables \ref{tab:tab1} and \ref{tab:tab2}. The parameter value represents the total number of neurons, excluding dropouts. Dropout refers to a layer that sets the inputs from the layer below to zero and eliminates a parameter value-sized fraction of them which are selected at random. By doing this, overtraining is avoided.


\begin{table}
  \caption{Setup of categorical LSTM-network}
  \label{tab:tab1}
  \begin{tabular}{|l|l|l|}

\hline
   \multicolumn{2}{c}{\textbf{LSTM-categorical} } \\

\hline

    Layer type & Parameter value & \vtop{\hbox{\strut Activation }\hbox{\strut function}}
\\
    \hline
    LSTM&   100&    tanh    \\
    \hline
    LSTM &    500&    tanh\\
    \hline
    Dropout&    .7&    -\\
    \hline
    LSTM&    700&    tanh\\
    \hline
    LSTM&    100&    tanh\\
    \hline
    Fully connected&    70&    Linear\\
    \hline
    tanh&    1&    softmax\\
    
  \hline
\end{tabular}
\end{table}

\begin{table}
	\caption{Setup of regressive LSTM-network}
	\label{tab:tab2}
	\begin{tabular}{|l|l|l|}
		
		\hline
		\multicolumn{2}{c}{\textbf{LSTM-regressive} } \\
		
		\hline
		
		Layer type & Parameter value & \vtop{\hbox{\strut Activation }\hbox{\strut function}}
		\\
		\hline
		LSTM&		100&		tanh\\
		\hline
		LSTM&		100&		tanh\\
		\hline
		Dropout&		.5&		-\\
		\hline
		LSTM&		50&		tanh\\
		\hline
		Fully connected&		50&		RelU\\
		\hline
		Fully connected&		70&		Linear\\
		\hline
		tanh&		1&		softmax\\
				\hline
	\end{tabular}
\end{table}

MSE was employed as the internal evaluation measure for the regressive network, and a mini-batch size of 30 was used to train the network. 2000 epochs were run on the network, but strangely, there was no sign of overtraining. However, after 600 epochs, advancement came to an end. The categorical network employed a minibatch size of 50 and categorical cross entropy as its evaluation measure. The ensuing experiments were stopped after 75 epochs due to overtraining, which was initially ran for a comparable amount of epochs. After 75 training epochs, the results for categorical are computed.

\textbf{LASSO}

The parameter in LASSO is $\gamma$, which was set to $\gamma  = 1$ which gives the algorithm a propensity to set less important features to zero.

\textbf{Random Forest}

Random forest created 10 trees from which it combined the results to create its output. No max depth was set, the number of possible nodes in one walk through the tree, instead the tree was determined to be large enough either because all leaves are pure or the smallest number of samples for split is reached, the smallest number was two. Purity calculations was done with Gini impurity for the classifier and MSE for regressor.

\textbf{SVC}

Sci-kit Learn's default settings were used to configure the Support Vector Classifier \cite{dutta2017depth}. All of them made use of the influx of failure data but turned them into classifications ranging from 1 to 6 depending on how much the inflow from the previous month had changed. as explained in Table \ref{tab:tab3} Regressive algorithm preprocessing was also carried out. Using a MinMax-scaler, the data was changed by rescaling the values to the [0,1] range for both the input and labeled output.

\begin{table}
	\caption{The classification of failure inflow depending on the difference in inflow between the current and previous month.}
	\label{tab:tab3}
	\begin{tabular}{|l|l|}
		\hline
		Category & Difference in inflow, x \\
		\hline
		6 &		$x > 30$\\
		\hline
		5 &		$15 < x \ge 30$\\
		\hline
		4 &		$5 < x \ge 15$\\
		\hline
		3 &		$-5 < x \ge 5$\\
		\hline
		2 &		$-15 < x \ge -5$\\
		\hline
		1 &		$x < -15$\\
		\hline
	\end{tabular}
\end{table}

\begin{table*}
	\caption{Performance from test 1 }
	\label{tab:tab4}
	\begin{tabular}{|l|l|l|l|l|l|l|l|}
		\hline
		Score& &	\multicolumn{1}{c}{Random Forest}& &LASSO&SVC&LSTM & \\
	\hline
	& & Categorical &
	Regressive& &  & Categorical &
	Regressive\\
			
		\hline
		MAE& Full set& 0.29& \textbf{0.04}& 0.06& 0.57& 2.11& 0.03\\
		\hline
		MAE& Test set& 1.36& 0.09& \textbf{0.06}& 1.8& 0.96&	0.11\\
		\hline
		R2&	Full set& 0.66&	0.91& 0.87&	0.32& -0.93& 0.92\\
		\hline
		R2& Test set& -0.41& 0.35& 0.72& 0&	-0.56& -0.13\\
		\hline
		MSE& Full set& 0.98& <0.01&	0.01& 1.90& 7.09&	<.01\\
		\hline
		MSE& Test set& 4.63& 0.01& 0.01& 6.06& 1.78& .02\\
		\hline
		f1& Full set& \textbf{0.88}& - & -& 0.79& \textbf{0.57}& - \\
		\hline
		f1& Test set& \textbf{0.42}& -&	-& 0.15& 0.22&	-\\
		\hline
		
	\end{tabular}
\end{table*}

The effectiveness of the described models when used on the test  1 dataset is displayed in Table \ref{tab:tab4}. It needs to be noted that the overall f1 score for the category LSTM model was 0.57. This would mean that the model almost mislabelled the data as it properly did not weigh the precision across the various labels. Compared to other category algorithms, this is worse. The random forest approach outperformed the second-best category algorithm on both the test set and the whole set, and it performed nearly twice as well. However, the test set performance in absolute terms was below 0.5, which indicates that there were more incorrect labels than right labels. The Regressive LSTM was comparable to other algorithms of the same type, unlike its categorical equivalent. This was true for both MAE and MSE, and based on the R2 score, it accounts for virtually all of the output's volatility. On the entire test set and on the test set, LASSO and random forest had the best performance.

\subsubsection{Addressing Research Question RQ 1:}

Which machine learning algorithm is best suited to predict the number of failures? 
LASSO algorithm performed the best overall throughout the test set. Any model should see a little increase in error from the training set to the test set. This was quite little when using the LASSO model, indicating that it did a good job of estimating the underlying data linkages. When utilizing both MAE and MSE metrics on the entire dataset, Random Forest and LSTM performed better, but their performance considerably increased when only looking at the error on the test sets. On the test set, the categorical models fared badly. There may be a high auto-correlation between the general trend of the inflow of difficulty reports and the category models' poor performance. In other words, while the inflow varies from month to month, it generally has a higher chance of declining if it has been declining in the past, and vice versa if it has been growing. The removal of portion of this association occurs once the input is transformed into categorical data. Tables \ref{tab:tab5} and \ref{tab:tab6}, which demonstrate the relative significance of the various data properties for the Random Forest Classifier and Regressor, respectively.

\begin{table}
	\caption{Random Forest Classifier: Top 5 data features}
	\label{tab:tab5}
	\begin{tabular}{|l|l|l|}
		\hline
		& Random Forest Classifier & \\
		\hline
		 & Feature (at time)& Importance \\
		\hline
		1& Mean development time (t-2)& 0.13\\
		\hline
		2& Mean development time (t-3)&	0.12\\
		\hline
		3& Mean development time (t-1)&	0.10\\
		\hline
		4& Number of releases (t)    & 0.08\\
		\hline
		5& Mean development time (t-4)&	0.08\\
		\hline
	\end{tabular}
\end{table}

\begin{table}
	\caption{Random Forest Regressor: Top 5 data features}
	\label{tab:tab6}
	\begin{tabular}{|l|l|l|}
		\hline
		& Random Forest Regressor & \\
		\hline
		& Feature (at time)& Importance \\
		\hline
		1& TR inflow(t-4) & 0.36\\
		\hline
		2& TR inflow(t-1)&	0.33\\
		\hline
		3& TR inflow(t-2)&	0.10\\
		\hline
		4& Mean development time (t-1)    & 0.05\\
		\hline
		5& TR inflow(t-3)&	0.05\\
		\hline
	\end{tabular}
\end{table}

\begin{figure*}[h!]
	\centering
	\includegraphics[scale=0.5]{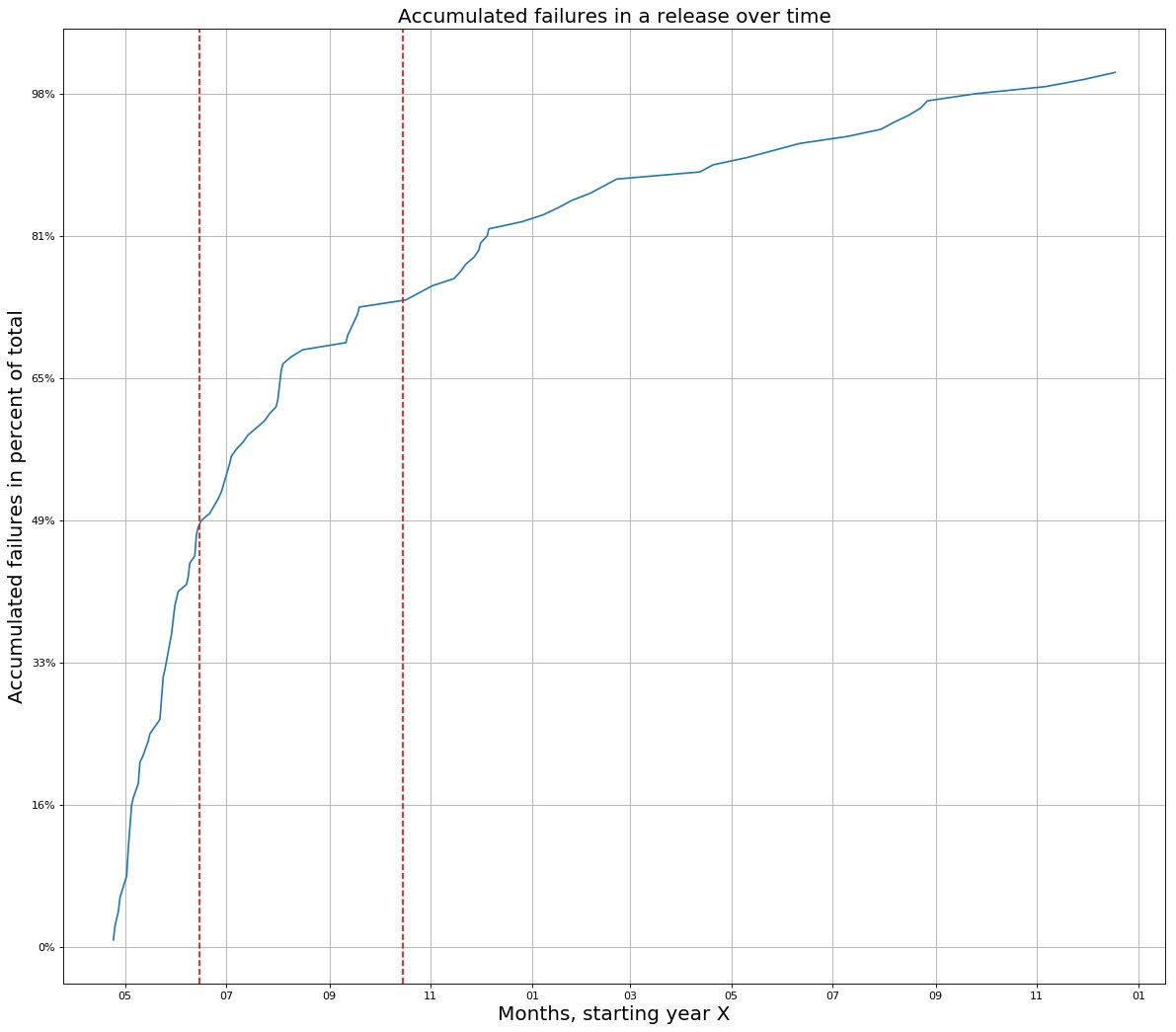}
	\caption{Graph showing cumulative inflow of trouble reports in percentage of total accumulated failures over time for a software made available in June of year X. First dashed vertical line is release date, the second dashed vertical line marks 4 months later.}
	\label{fig:fig4}
\end{figure*}

The decision to abandon the categorical LSTM model was reached in light of the findings. However, the other classifiers were preserved. This was due to the fact that they relied less on the TR data and could be utilized to determine the significance of the new data in test 3. In addition, it was decided to integrate a CNN model in the third test. There is evidence that they are more resilient when transferring to different situations, in part because of the speedier training. Which may be pertinent to the business, for instance, if the network is applied to a different product inside the business.
The choice was taken to base the time series since difficulty reports are identified by the release in which they were found. The total number of failures found in a software may span more than a year since one version may be in use for a considerable amount of time before a customer chooses to utilize a subsequent one, as shown in Figure \ref{fig:fig4}. An analysis of this behavior reveals that, on average, 63\% of all defects are found before the product is released. 58\% of failures that are found when the version is in use are found during the first four months following release.

\subsection{Test 2 Setup}
Data from the software components that made up were examined in Test 2. The number of lines of code that had been added, modified, or deleted as part of the development of a new software feature was disclosed in this data. It was also possible to see which systems or subsystems each code modification affected. By looking at the date on which the final modification in the particular software feature was made, it was possible to identify which version the feature was a part of. As a result, clustering was performed on the software features, as determined by the system change attributable to the feature; the metrics are described below.

In test 2, the clustering was performed in two distinct ways. The two versions represented the system changes differently, using distinct software features and clustering techniques. The first dealt with systems. Each modification to a system was counted either by counting the number of impacted files or the number of lines of code inside that system. The clustering methods were then applied to both versions after pre-processing with the maxmin-, robust-, quantile uniform-, and quantile normal-scalers. The second version was performed at the level of subsystems, where a comparable computation was made. 

The clustering methods were then applied to both versions after pre-processing with the maxmin-, robust-, quantile uniform-, and quantile normal-scalers. Grid-search was used to choose the parameters for DBSCAN with m = [5,20] and range = [0.1,1.5]. In order to prevent providing too much data to test 2, the best solutions were then chosen by balancing the silhouette score with a lesser number of clusters, 30.
The size of the clusters varied greatly. This would have resulted in very little time series structured input data for test 2. Although there are more reliable approaches to utilize for regression issues with sparse data \cite{zhong2004analyzing}, it was determined that this may be troublesome for the models and might not be accurate in representing the situation as it actually is. This was due to the fact that, although if some software features may undoubtedly be exceptions, it appeared implausible that the majority of all features generated would be the same. Due to this, the performance of the clustering algorithm was evaluated using the silhouette score and cluster distribution.

\begin{table*}[h!]
	\caption{Results from clustering software features on system level}
	\label{tab:tab7}
	\begin{tabular}{|l|l|l|l|l|}
		\hline 
				\multicolumn{5}{c}{\textbf{System} } \\
		\hline
	& Number of files & & 
	
	\multicolumn{2}{c}{Number of lines of code} \\
	\hline
	K-means & Silhouette Score&	Largest cluster& Silhouette Score&	Largest cluster\\
	\hline
MinMax& 0.929& 98,2\%& 0.985& 99,3\\
		\hline
Robust v1& 0.424& 66,9\%& 0.854& 96,1\\
		\hline
Robust v2& 0.399& 60,1\%& 0.906& 96,6\\
		\hline
Quantile Uniform& 0.537& 22,8\%& 0.481& 25,3\\
		\hline
Quantile Normal& 0.522& 22,6\%& 0.486& 25,3\\
		\hline
DBSCAN& & Clusters & &  Clusters\\
		\hline
MinMax& 0.993& 2& 0.995& 2\\
		\hline
Robust v1& 0.741& 25& -0.205& 2\\
\hline
Robust v2& 0.741& 25& -0.205& 2\\
\hline
Quantile Uniform& 0.771& 28& 0.677& 28\\
\hline
Quantile Normal& 0.776&	28&	0.724& 29\\
		\hline
	\end{tabular}
\end{table*}

\subsubsection{Addressing Research Question RQ 2:}

How can machine learning be used to group software attributes into clusters to help anticipate failure?
On the same datasets, K-means and DBSCAN were tested and contrasted. First, a systems level representation of the program features was used. One feature might have an impact on several systems, thus two distinct metrics were employed to gauge how much each system was affected. Table \ref{tab:tab7} provides the results at the systems level, whereas Table \ref{tab:tab8} lists the results at the subsystem level. The findings seemed to imply that there is quite a lot of noise in the dataset, the results appeared to suggest that there is quite a bit of noise in the dataset, which can make estimating fault-proneness more challenging.

\begin{table*}[h!]
	\caption{Results from clustering software features on subsystem level}
	\label{tab:tab8}
	\begin{tabular}{|l|l|l|l|l|}
		\hline 
		\multicolumn{5}{c}{\textbf{Subsystem} } \\
		\hline
		& Number of files & & 
		
		\multicolumn{2}{c}{Number of lines of code} \\
		\hline
		K-means & Silhouette Score&	Largest cluster& Silhouette Score&	Clusters\\
		\hline
	MinMax& 0.89& 97,0\%& 0.95& 10\\
	\hline
	Robust v1& 0.13& 79,2\%& 0.87& 10\\
	\hline
	Robust v2& 0.62& 88,6\%& 0.87& 10\\
	\hline
	Quantile Uniform& 0.27& 34,7\%& 0.26& 10\\
	\hline
	Quantile Normal& 0.25& 43,3\%& 0.25& 10\\
	\hline
	DBSCAN& & Clusters & &\\
	\hline
	MinMax& 0.98& 2& 0.99& 2\\
	\hline
	Robust v1& 0.47& 47& -0.55& 16\\
	\hline
	Robust v2& 0.47& 47& -0.55& 16\\
	\hline
	Quantile Uniform& 0.48& 2& 0.49& 2\\
	\hline
	Quantile Normal& 0.36& 20& 0.34& 20\\
\hline
\end{tabular}
\end{table*}

Depending on the pre-processor, DBSCAN can identify noise and outliers, which make up between 5 and 15\% of the data at the system level. It was concluded that the DBSCAN algorithm's clusters may be more suited than the k-means ones for this reason.
\textbf{RQ 2.1:} What data regarding the software features is best used for clustering?
The feature data describing the quantities of files modified on a systems level was found to be the best for clustering on because the clustering performance was evaluated on the dual criteria of silhouette score and cluster distribution.

\subsection{Test 3 Setup}

The data used in test 2 was an amalgamation of the data used in test 1 and that created in test 2. For the neural networks the layout was determined through a grid search, the result of which are shown in Tables \ref{tab:tab9} and \ref{tab:tab10}, for the LSTM and CNN, respectively. The other algorithms had the same setups as in test 1. The data generated in test 2 was used and the output was increased to also consider how important a failure was considered. Ranging from 1, extremely important, 2, less important but still important enough to stop distribution of whatever feature caused it. While 3 and 4 were critical enough to be fixed, 5 was viewed as essentially unimportant.

\begin{table}[h!]
	\caption{Settings of the LSTM-network used}
	\label{tab:tab9}
	\begin{tabular}{|l|l|l|}
		\hline 
		\multicolumn{3}{c}{\textbf{LSTM} } \\
		\hline
		Layer type& Parameter value&  \vtop{\hbox{\strut Activation }\hbox{\strut function}}\\
		\hline
		Gaussian Noise&	0.05& -\\
		\hline
		LSTM& 200& tanh\\
		\hline
		Dropout&.3&	-\\
		\hline
		LSTM& 150& tanh\\
		\hline
		LSTM& 200& tanh\\
		\hline
		Fully-connected& 70& Linear\\
		\hline
		Fully-connected& 6& Sigmoid\\
		\hline
	\end{tabular}
\end{table}

\begin{table}[h!]
	\caption{Settings of the CNN-network used}
	\label{tab:tab10}
	\begin{tabular}{|l|l|l|}
		\hline 
		\multicolumn{3}{c}{\textbf{CNN} } \\
		\hline
		Layer type& Parameter value&  \vtop{\hbox{\strut Activation }\hbox{\strut function}}\\
		\hline
		Gaussian Noise&	0.05& -\\
		\hline
		Conv1D&	64, 2& tanh\\
		\hline
		RelU& .3& -\\
		\hline
		Dropout& .3& -\\
		\hline
		MaxPooling1D& 200& tanh\\
		\hline
		Fully connected& 70& Linear\\
		\hline
		tanh& 6& Sigmoid\\
		\hline
	\end{tabular}
\end{table}

\subsubsection{Results and Addressing Research Questions 2.2:}

Are certain clusters more likely to contribute to failures in the software, and if so can they be used to increase performance of the prediction models?

\begin{table*}
	\caption{Data feature relevance for inflow of failures}
	\label{tab:tab11}
	\begin{tabular}{|l|l|l|l|l|l|l|l|l|l|}

		\hline
		\multicolumn{2}{c}{Feature rank} & 1 & 2 & 3 & 4 & 5 & 6 & 7 & 8 \\
		\hline
		
		Correlation & Data Feature &\textbf{-1} &-1 &\textbf{4}& 4& \textbf{0}& 3&9 &2 \\
		\hline
		Correlation&time step&-1&-2&-1&-1&-1&-1&-2&-1 \\
		\hline
		Correlation&Importance&-0,69&-0,65&-0,59&-0,56&-0,55&-0,52&-0,52&-0,51\\
		\hline
		LASSO& Data Feature&20&	7&Med&Long&	20& 17& 22& 13\\
		\hline
		LASSO& time step&-2& 0&	-2&	-1&	0&-2&0&	-1\\
		\hline
		LASSO& Importance& 0,25& 0,24& 0,22& 0,21& 0,18&0,13&0,13&0,12\\
		\hline
		Random Forest&Data Feature& \textbf{4}&\textbf{0}&\textbf{-1}&	12&	4&12&-1&2\\
		\hline
		Random Forest&time step&-1&-1&-1&-1&0&-2&-2&-1\\
		\hline
		Random Forest&Importance&0,13&0.11&	0.11&0,07&0,07&0,06&0,06&0,05\\
		\hline
	\end{tabular}
\end{table*}

\begin{table*}
	\caption{Data feature relevance for importance class 2}
	\label{tab:tab12}
	\begin{tabular}{|l|l|l|l|l|l|l|l|l|l|}
		
		\hline
		\multicolumn{2}{c}{Feature rank} & 1 & 2 & 3 & 4 & 5 & 6 & 7 & 8 \\
		\hline
		
		Correlation & Data Feature &\textbf{-1}&-1&9&\textbf{4}&3&3&4&9 \\
		\hline
		Correlation&time step&-1&-2&-2&-2&-1&-2&-1&-1\\
		\hline
		Correlation&Importance&-0,60&-0,55&-0,48&-0,45&-0,43&-0,41&-0,40&-0,39\\
		\hline
		LASSO& Data Feature&-0,60&-0,55&-0,48&-0,45&-0,43&-0,41&-0,40&-0,39\\
		\hline
		LASSO& time step&0&-2&0&0&-1&0&-2&-2\\
		\hline
		LASSO& Importance& 1,4&1,0&0,81&0,79&0,74&0,58&0,48&0,45\\
		\hline
		Random Forest&Data Feature&\textbf{-1}&-1&9&12&Med&Med&Short&26\\
		\hline
		Random Forest&time step&-2&-1&-2&-1&-1&0&-1&0\\
		\hline
		Random Forest&Importance&0,18&0,099&0.083&0,072&0,057&0,041&0,037&0,036\\
		\hline
		\end{tabular}
\end{table*}

\begin{table*}
	\caption{Data feature relevance for inflow of failures}
	\label{tab:tab13}
	\begin{tabular}{|l|l|l|l|l|l|l|l|l|l|}
		
		\hline
		\multicolumn{2}{c}{Feature rank} & 1 & 2 & 3 & 4 & 5 & 6 & 7 & 8 \\
		\hline
	Correlation & Data Feature &-1&-1&4&4&0&0&3&9\\
	\hline
	Correlation&time step&-2&-1&-1&-2&-1&-2&-1&-2\\
	\hline
	Correlation&Importance&-0,62&-0,61&-0,58&-0,56&-0,53&-0,52&-0,51&-0,51\\
	\hline
		LASSO& Data Feature&12&Long&17&20&26&14&24&Med\\
		\hline
		LASSO& time step&0&-1&-1&-2&-2&-2&-2&0\\
		\hline
		LASSO& Importance& 0,63&0,62&0,53&0,47&0,15&0,14&0,14&0,14	\\
		\hline
		Random Forest&Feature number& \textbf{4}& \textbf{-1}&12&\textbf{0}&0&5&0&17\\
		\hline
		Random Forest&time step&-1&-1&-1&-1&0&-2&-2&-1\\
		\hline
		Random Forest&Importance&0,23&0.070&0.069&0,063&0,059&0,050&0,031&0,028\\
		\hline		
	\end{tabular}
\end{table*}

Tables \ref{tab:tab11}, \ref{tab:tab12}, and \ref{tab:tab13} show what data features the algorithms use to predict the inflow of failures, and the inflow of failures of importance class 2 and 4, respectively. The tables display the most significant data elements for the overall inflow of difficulty reports as well as the inflow of reports with significance classes 2 and 4. Several software feature groups: -1 (noise), 0, and 4 are shared by Random Forest and correlation, although notably neither 0 nor 4 seem significant to Random Forest for significance class 2. However, LASSO makes use of other. There is some unpredictability in noisy data, which is sometimes modeled using a normal distribution. However, in the instance of DBSCAN, the noisy data is isolated from the other datapoints because it is unclear whether they are a part of a cluster or not. Since they are distinct from one another, it makes sense that the software features that are notably different from other software features that have already been produced are going to be more complicated. The other clusters that appear repeatedly in both Random Forest and Correlation may have some innate characteristics that make them more prone to failure.

\begin{table*}
	\caption{Performance from test 2}
	\label{tab:tab14}
	\begin{tabular}{|l|l|l|l|l|l|l|l|}
		\hline
		Score& &	\multicolumn{1}{c}{Random Forest}& &Lasso&
		SVC&LSTM&CNN
		\\
		\hline
		& & Categorical &	Regressive& &  &  &\\
		
		\hline
		MAE& Full set& 0.21 &\textbf{0.06}&	0.04&0.18&\textbf{0.05}& 0.05\\
				\hline
		MAE& Test set&1.0&\textbf{0.11}&0.22&0.809&\textbf{ 0.11}&	0.16\\
		\hline
		R2&	Full set&0.76&0.86&	0.68& 	0.82&0.86&	0.80\\
		
		\hline
		R2& Test set&-0.01&	0.31& 	-1.17&0.0&-4.46&-2.91\\
		\hline
		MSE& Full set&0.39& 0.01&0.018&0.27&	0.01&0.01\\

		\hline
		MSE& Test set& 	1.89&0.02&0.091&1.33&0.03&0.04\\
		\hline
		f1& Full set& \textbf{0.86}&	-&-&0.86&-&	-\\
			\hline
		f1& Test set& 	\textbf{0.18}&-&	-&0.17&-&-\\
		\hline
		
	\end{tabular}
\end{table*}

If understanding certain clusters that are more likely to result in failures will help the models, what is the second component of research question 3? To respond, Table \ref{tab:tab14} demonstrates that there was no discernible advancement between the models in test s 1 and 3. As a result, the LSTM network only surpasses random forest on the test set and becomes the top algorithm overall. However, more interestingly, some actually became worse. For example, the category Random Forest model's f1-score on the test set drastically declined from 0.42 to 0.18, despite being similar for the entire collection. Additionally, LASSO fared worse on the regressive issues than in test 1.The three distinct methods for calculating feature significance only barely agree with one another when it comes to the first section of the study question. Therefore, even while some clusters do seem to be more prone to have failures, the models do not appear to benefit.

\begin{figure*}[h!]
	\centering
	\includegraphics[scale=0.5]{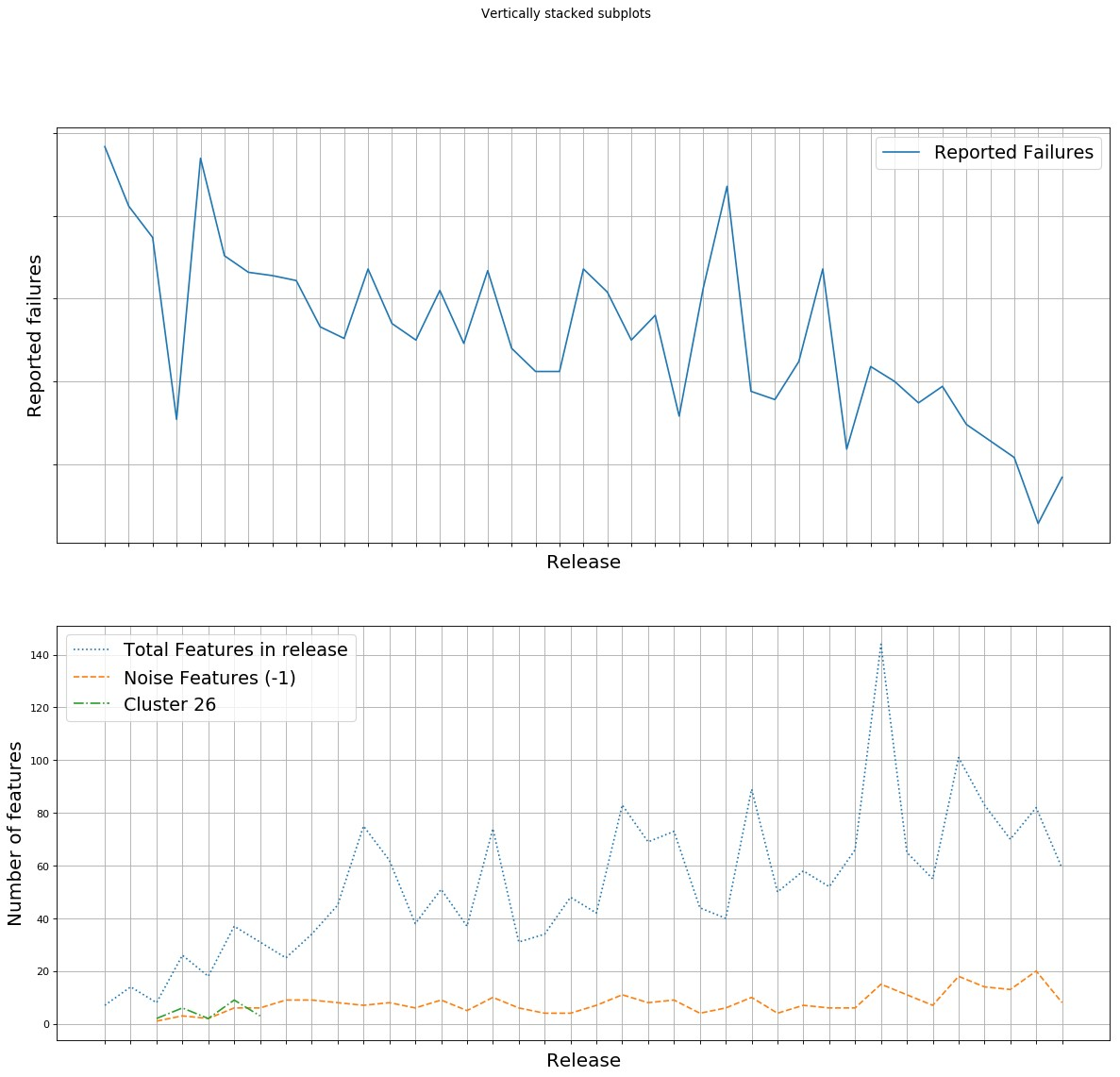}
	\caption{Top graph shows of accumulated failures per release (actual numbers removed due to company privacy concerns), bottom graph shows total software features delivered, and software features of type -1 and 26 per release included in the study.}
	\label{fig:fig5}
\end{figure*}


The top graph in Figure \ref{fig:fig5} displays the quantity of failures each release. The y-axis lacks a scale due to privacy issues with the corporation. However, it is still obvious that the number of bug reports per release is declining, whereas the trend for developed software features is the opposite. Since the correlation is negative, it is challenging to determine which software characteristics mostly cause an increase in failures using a linear model. Instead, it would seem that greater software features should result in fewer difficulty reports under this paradigm. DBSCAN labels noise as cluster 1, which has a significant negative correlation and is one of the more crucial data properties in Random Forest. Therefore, forecasts can be used, but the inclusion of noisy software features in a release may not be taken to indicate a higher risk of failure.

\section{Conclusion}
This paper examined the ability of specific machine learning methods to predict cumulative failure. Additionally, it sought to comprehend how data-based clustering of software characteristics might be accomplished. This clustering was done to determine which clusters were more prone to errors. The cumulative failure prediction was further extended using clustering. Initial cumulative failure prediction, clustering, and extended cumulative failure prediction were the three phases that made up the three tests. What data factors most significantly influenced the cumulative failure prediction was of further interest. Some of the techniques used to address this problem were white-box, enabling understanding of how the data was weighted to produce the results. Five different supervised machine learning techniques were applied to forecast cumulative failure. There were two possible approaches to set up the issue: as a classifier problem, where the number of failures was compared to the number of failures in the past and then categorized based on the amount of the increase or reduction. Additionally, the inflow was to be predicted as a regression issue, scaled to the methods, with the goal of doing so. Regressors were correct with regard to both the whole set and test set (error 10\%), despite the fact that they were applied to distinct data sets and shouldn't be directly compared. The classifiers did okay on the entire set (f1-score of 88\%), with random forest having the best performance among the classifiers.
On the complete set and test set, respectively, random forest and LASSO had the best test 1 results. The test set performance of LASSO in test 2 declined, nevertheless, the LSTM-network, which used grid search to fine-tune its parameters, delivered the best results.
Due to the lack of certain data, clustering of the software features, which was performed in test 2, cannot be evaluated using an absolute metric. Since it was obvious that the data had some noise, DBSCAN was chosen as the preferable option since it can tolerate noise and classify outliers into their own group. Data on the quantities of files altered on a systems level were utilized to obtain both good clustering results, as measured by the outlined score, and a suitable distribution of the number and size of the clusters.
Test 2 expanded the application and new data to the initial issue and looked at whether specific groups of software characteristics were more likely to increase the chance of failures. There were some parallels between random forest and the link between the amount of features in a particular cluster and the cumulative failure. However, none of the clusters they suggested were the same as those suggested by LASSO. The fact that the overall pattern of failures is a decrease in the number of failures every release while the overall number of features is growing further obscured these conclusions. The overall trends indicate that it is not only the complexity of a certain feature that determines likelihood of failure but also other organizational decisions that affect the work performed. As is indicated by the fact that despite more features being created they result in an overall net-decrease of failures.


\clearpage
\bibliographystyle{ACM-Reference-Format}
\bibliography{./sample-base}

\end{document}